\documentclass[sigconf]{acmart}
\AtBeginDocument{%
  }

\setcopyright{acmlicensed}
\copyrightyear{2025}
\acmYear{2025}
\acmDOI{XXXXXXX.XXXXXXX}
\acmConference[PEARC '25]{Make sure to enter the correct
  conference title from your rights confirmation email}{June 03--05,
  2025}{Woodstock, NY}
\acmISBN{978-1-4503-XXXX-X/2018/06}

\usepackage{multirow}
\usepackage{subcaption}

\begin{document}

\title{A Pilot Study on Tunable Precision Emulation via Automatic BLAS Offloading}

\author{Hang Liu}
\email{hliu@tacc.utexas.edu}
\orcid{0000-0002-3486-7863}
\author{Junjie Li}
\email{nicejunjie@gmail.com}
\orcid{0000-0002-1051-5927}
\author{Yinzhi Wang}
\email{iwang@tacc.utexas.edu}
\orcid{0000-0001-8505-0223}
\affiliation{%
  \institution{Texas Advanced Computing Center, The University of Texas at Austin}
  \city{Austin}
  \state{Texas}
  \country{USA}
}

\begin{abstract}
This study explores the use of automatic BLAS offloading and INT8-based emulation for accelerating traditional HPC workloads on modern GPU architectures. Through the use of low-bitwidth integer units and cache-coherent Unified Memory Architecture, we emulate double-precision matrix multiplications in the MuST application without code changes. We find that accuracy depends on both arithmetic precision and the properties of the operator, which can be dealt with through tunable precision emulation. Unlike traditional mixed-precision approaches, this method preserves original algorithms while optimizing hardware utilization. We showcases the potential of improving accuracy and performance at the same time. This work highlights the potential of AI-driven hardware to transform HPC, advocating for adaptive precision strategies in future scientific computing.
\end{abstract}




\maketitle

\section{Introduction}
Rapid advancement in artificial intelligence (AI) has triggered the development of specialized hardware, such as NVIDIA GPUs with Tensor Cores, AMD GPUs with Matrix Cores and various TPUs, to accelerate machine learning computations with high-throughput and low-precision arithmetic units\cite{abdelfattah_survey_2021}. 
These processors take advantage of low/mixed-precision floating-point formats (e.g., FP16, BF16) and low-bitwidth integer units (e.g., INT8, INT4) to increase performance, reduce power consumption, and minimize memory footprint, as in the case of technologies like NVIDIA's DP4A instruction and integer Tensor Cores \cite{horowitz_11_2014}. While these architecture innovations have had a dramatic impact on AI training and inference performance, their potential to accelerate traditional HPC workloads that were historically based on CPUs and double-precision floating point (FP64) computations for scientific simulation, numerical modeling, and other compute-intensive tasks, remains vast yet not fully understood, requiring further research and development to unleash.

FP64 has been the norm in traditional HPC, ensuring the numerical precision required for simulations like climate modeling, quantum chemistry simulation, and computational fluid dynamics. However, not all of the calculations in science demand that level of precision; lower precision may be enough in some cases without compromising result validity. Meanwhile, certain special computations, such as in quantum chemistry, can require higher precision than FP64. This level of precision needed provides an opportunity to rethink computational strategies in HPC. By offloading suitable workloads on GPUs and leveraging reduced-bit-width integer types like INT8, one can emulate FP64 computations at acceptable precision and utilize fewer computational resources while achieving a huge performance gain. For example, NVIDIA Tensor Cores are capable of performing INT8 matrix multiplications with INT32 accumulation at theoretical maximum performance rates 2 to 4 times higher than their floating point counterparts \cite{ootomo_dgemm_2024}. Similarly, more recent processors such as Google TPUs and Intel AMX-INT8 demonstrate that integer acceleration for intensive matrix workloads is possible \cite{jouppi_datacenter_2017}.

The limitation of using GPU acceleration in traditional HPC workflows lies in the challenge of retrofitting legacy CPU-based applications, which are often deeply embedded within FP64-based numerical libraries and complex workflows. Manual porting is labor-intensive and not a viable choice for the majority of scientific communities, and existing automatic offloading capabilities have previously offered relatively modest performance improvement due to inefficiencies in traditional CPU-GPU memory architectures. However, recent innovations such as NVIDIA's Grace-Hopper superchip's Unified Memory Architecture (UMA) and AMD's MI300A APU enable cache-coherent memory access between the GPU and CPU, ease data handling, and allow more transparent offloading strategies \cite{li_automatic_2024}. All these advances foreshadow a paradigm shift for HPC, away from CPU-centric FP64 computations toward adaptive precision strategies that leverage AI-driven hardware for both performance and accuracy. This crossroads of hardware innovation and modern memory architectures motivates our pilot study: examining whether automatically offloading Basic Linear Algebra Subprograms (BLAS) operations to GPUs, with tunable precision emulation using INT8, can maximize the efficiency of HPC workloads. By tuning the precision when needed, through the use of the integer representations, we can improve performance as well as the potential for better numerical precision for scientific computing.

This work examines the application of automatic BLAS offloading \cite{li_automatic_2024} together with the Ozaki scheme on Integer Matrix Multiplication Units (IMMU) for the emulation of matrix multiplications in INT8 \cite{ootomo_dgemm_2024,uchino_performance_2025}. We experiment with a CPU application, MuST \cite{must_1995}, through GPU offloading and emulation across various precision levels. The Ozaki scheme provides a general framework that we adapt here to exploit the accuracy of the application using the INT8 arithmetic units of GPUs. Our work demonstrates that MuST's accuracy is not only determined by the arithmetic precisions, but also influenced by the operator's analytic properties on its energy domain, revealing an accuracy sensitivity that can be addressed through tunable precision. This showcases the potential of improving accuracy without trading off performance by the automatic offloading and tunable precision emulation on modern GPU architectures. 

\section{Related Work}
\subsection{Previous Work on Automatic BLAS Offload}
Several attempts have been made to automatically offload BLAS calls from CPUs to GPUs with the rise of GPU adoption. Cray LIBSCI\_ACC \cite{cray-libsci-acc, cray-libsci-acc-slide} pioneered the approach on the Titan supercomputer and supports offloading selected BLAS, LAPACK, and ScaLAPACK routines when the accelerated module is used. IBM's ESSL \cite{ibm-essl-offload} enables automatic offload of BLAS, LAPACK, and FFTW calls when binary is linked to libesslsmpcuda. NVIDIA's NVBLAS \cite{nvidia-nvblas} used LD\_PRELOAD to divert CPU BLAS calls to the GPU without relinking. 
These tools automate data movement and offload decisions. 
However, they  are all designed for the conventional GPU architectures where data transfer between CPU and GPU is mandatory for every offloaded BLAS call, resulting in high overheads and heavy loss of performance in practical HPC applications \cite{li_automatic_2024, li_automatic_2025}.

A significant advancement comes from Li et al. \cite{li_automatic_2024,li_automatic_2025}, who developed a tool for automatic BLAS offloading on NVIDIA Grace-Hopper superchip. Leveraging the cache coherency in UMA, the library intercepts compute intensive level-3 BLAS calls using a light-weight, trampoline-based Dynamic Binary Instrumentation (DBI) approach \cite{wang_peak_2023}, and offers three data movement strategies tailored for the new GPU architecture: traditional CPU-GPU memory copies, cahce-coherent unified memory access, and an optimal  migration scheme moving data upon GPU first use. Tested on real HPC applications, it achieves substantial performance gains without code changes, overcoming prior tools’ limitations and setting a new standard for BLAS offloading in HPC.

\subsection{DGEMM Emulation on Integer Matrix Multiplication Unit}
The study of floating point emulation isn't new. 
The most widely known emulation is emulating quad-precision using double-precision \cite{hida_algorithms_2001} for those scientific applications that mandates precision beyond FP64. 
In recent years, a series of paper on the Ozaki scheme\cite{ozaki_error-free_2012, ozaki_generalization_2013, ozaki_tc_2020} were published exploring generic error-free emulation schemes for matrix multiplications using lower-precision variable types such as FP16. 
Such scheme splits matrices into lower-precision slices, and accumulate result at higher-precision through a series of matrix multiplication operations on the lower-precision ones. 
A striking development made recently was the Ozaki Scheme on Integer Matrix Multiplication Unit (ozIMMU) by Ootomo et al.\cite{ootomo_dgemm_2024} and later enhanced by Uchino et al.\cite{uchino_performance_2025}, which implements the Ozaki scheme using INT8 Tensor Cores for emulating GEneral Matrix Multiplication (GEMM), avoiding rounding error in FP16 arithmetic and further improved accuracy and efficiency. 
The overall GEMM precision in OzIMMU can be controlled by adjusting the number of splits, and such  tunability is critical to our work, enabling efficient [D,Z]GEMM emulation on GPU for HPC workloads.   

It is worth noting that the term "emulation" is often incorrectly used to refer mix-precision, although the two approches are fundamentally different in both principle and applicability.  
In typical mix-precision methods, numerical algorithms are tweaked to adopt lower-precision in the more precision-tolerant components. 
Such approach is highly efficient as it avoids the overhead with emulation. However, it usually requires algorithmic modifications and is only applicable to specific, limited problems.
There are several recent reviews of the topic \cite{mix-prec-review1,abdelfattah_survey_2021,mix-prec-review3}.  
While being distinct, our study of precision tuning via emulation may offer a relatively straightforward path for adopting mixed-precision in complex codes or algorithms.

\section{Experiment}
\subsection{Implementation}
Our experiments use two open-source tools: ozIMMU \cite{tsuki_enp1s0ozimmu_2025} and SCILIB-Accel\cite{scilib-accel}. OzIMMU redirects cuBLAS function calls to its IMMU-based implementation using the dynamic linker, enabling INT8-based DGEMM and ZGEMM emulation. SCILIB-Accel, utilizing the PEAK profiler framework \cite{wang_peak_2023}, employs DBI to transparently intercept BLAS calls and automatically offload them to GPUs through calling cuBLAS. The transparent DBI implementation by design allows both tools to operate collaboratively with a single LD\_PRELOAD to achieve automatic offloading with tunable precision emulation on any BLAS-heavy CPU programs without code changes or recompiling. Note that we also applied the optimized ozIMMU\_H method \cite{uchino_performance_2025} to improve performance. In order to run a CPU code on the GPU's INT8 tensor cores, one has to set \verb$LD_PRELOAD=scilib-accel/scilib-dbi.so:ozIMMU_H/libozimmu.so$.
GEMM is computed using the mode set by \verb$OZIMMU_COMPUTE_MODE=mode$, with ozIMMU supporting modes like dgemm (native FP64 cuBLAS) and fp64\_int8\_3 to fp64\_int8\_18 (INT8 with split numbers 3 to 18).

\subsection{Experiments with MuST}
The case we choose to test the precision under different INT8 modes in ozIMMU is the MT u56 benchmark case in MuST distribution. Since the major solver in this LSMS case is LU based matrix invert, its zgemm intensity makes it a perfect target to test the accuracy under various INT8 modes in ozIMMU. The CPU version of MuST is built and run by NVHPC/24.7 with the nvhpc-hpcx-cuda12 module loaded on one Vista \cite{vista} node. We run it with INT8 modes from fp64\_int8\_3 to fp64\_int8\_9 as well as the dgemm mode. To compare the accuracy of the results from INT8 modes and dgemm mode by relative error the MuST output quantity Int[Z*Tau*Z-Z*J] is used. It is Green's function integrated over space as defined on the energy contour. We denote it as $G(z)$ for brevity. The total energy and Fermi energy are the results from integrating the Green function over a set of energy points. Therefore $G(z)$ is more critical to accuracy test. All $G(z)$ outputs for the atom id=1 in MT on each energy point and at each iteration are sorted out. The relative error of INT8 modes to dgemm mode is evaluated as $|G(z)_{dgemm}-G(z)_{int8}|/|G(z)_{dgemm}|$ to its real and imaginary parts respectively and the max\_real and max\_imag are the maximum relative error for each on all $z$. This offers a panoramic view to the accuracies of $G(z)$ from all ozIMMU INT8 modes through the whole calculation of this case.

\section{Discussion}
The maximum relative error results are in Table \ref{tab:split_modes}. As expected, the INT8 modes of split number 3 and 4, which are close to FP32, are likely not enough for acceptable accuracy in realistic calculation. When split numbers are 5 and 6, the accuracy is exponentially improved. The max\_real and max\_imag of $G(z)$ are down to $10^{-9}$ or $10^{-8}$, the total energy and Fermi energy are converged in term of INT8 split numbers and are equivalent to the results by FP64. This is an acceptable accuracy if only ground state and total energy are concerned. When split numbers are 7 or 8, the accuracy is equivalent to FP64. To our experience, the same level of the relative error occurs when building FP64-native codes by different compilation options and/or on different systems. The split number 9 seems exceeding the FP64 accuracy, but it may not be reflected in the result since only GEMM workload is running with higher accuracy, other calculations are still with FP64 accuracy.

\begin{table*}[h!]
\centering
\caption{Impact of Split Numbers on Accuracy across Iterations}
\vspace{-4mm}
\resizebox{\textwidth}{!}{
\begin{tabular}{l|rrrr|rrrr|rrrr}
\hline
             & \multicolumn{4}{c|}{iteration 1}        & \multicolumn{4}{c|}{iteration 2}        & \multicolumn{4}{c}{iteration 3}        \\
mode & max\_real   & max\_imag & Etot & Efermi & max\_real   & max\_imag & Etot & Efermi & max\_real   & max\_imag & Etot & Efermi \\
\hline
dgemm        & \multicolumn{1}{l}{}            & \multicolumn{1}{l}{}          & -0.550930                 & 0.72503                    & \multicolumn{1}{l}{}            & \multicolumn{1}{l}{}          & 79.384613                & 0.73449                    & \multicolumn{1}{l}{}            & \multicolumn{1}{l}{}          & -0.705320                 & 0.72651                    \\
int8\_3      & 3.85E-02                        & 2.63E-03                      & -0.551429                & 0.72503                    & 4.77E-03                        & 1.07E-02                      & 79.259925                & 0.73439                    & 3.28E-03                        & 4.15E-03                      & -0.705216                & 0.72646                    \\
int8\_4      & 7.94E-04                        & 9.07E-05                      & -0.550946                & 0.72503                    & 1.40E-04                        & 1.53E-04                      & 79.381569                & 0.73449                    & 1.25E-04                        & 7.23E-05                      & -0.705329                & 0.72651                    \\
int8\_5      & 8.36E-06                        & 8.84E-07                      & -0.550930                 & 0.72503                    & 1.92E-06                        & 2.10E-06                      & 79.384565                & 0.73449                    & 1.69E-06                        & 1.31E-06                      & -0.705320                 & 0.72651                    \\
int8\_6      & 8.25E-08                        & 3.96E-09                      & -0.550930                 & 0.72503                    & 2.51E-08                        & 2.69E-08                      & 79.384612                & 0.73449                    & 7.95E-09                        & 1.31E-08                      & -0.705320                 & 0.72651                    \\
int8\_7      & 4.20E-10                        & 8.64E-11                      & -0.550930                 & 0.72503                    & 2.38E-10                        & 2.45E-10                      & 79.384613                & 0.73449                    & 7.44E-11                        & 9.56E-11                      & -0.705320                 & 0.72651                    \\
int8\_8      & 4.90E-12                        & 1.54E-12                      & -0.550930                 & 0.72503                    & 1.74E-12                        & 2.22E-12                      & 79.384613                & 0.73449                    & 5.70E-13                        & 1.25E-12                      & -0.705320                 & 0.72651                    \\
int8\_9      & 1.15E-13                        & 3.54E-14                      & -0.550930                 & 0.72503                    & 9.82E-14                        & 7.37E-14                      & 79.384613                & 0.73449                    & 5.96E-14                        & 2.28E-13                      & -0.705320                 & 0.72651                   \\
\hline
\end{tabular}
}
\label{tab:split_modes}
\end{table*}

\begin{figure}[h!]
  \centering
  \begin{subfigure}[b]{0.42\textwidth}
    \includegraphics[width=\linewidth]{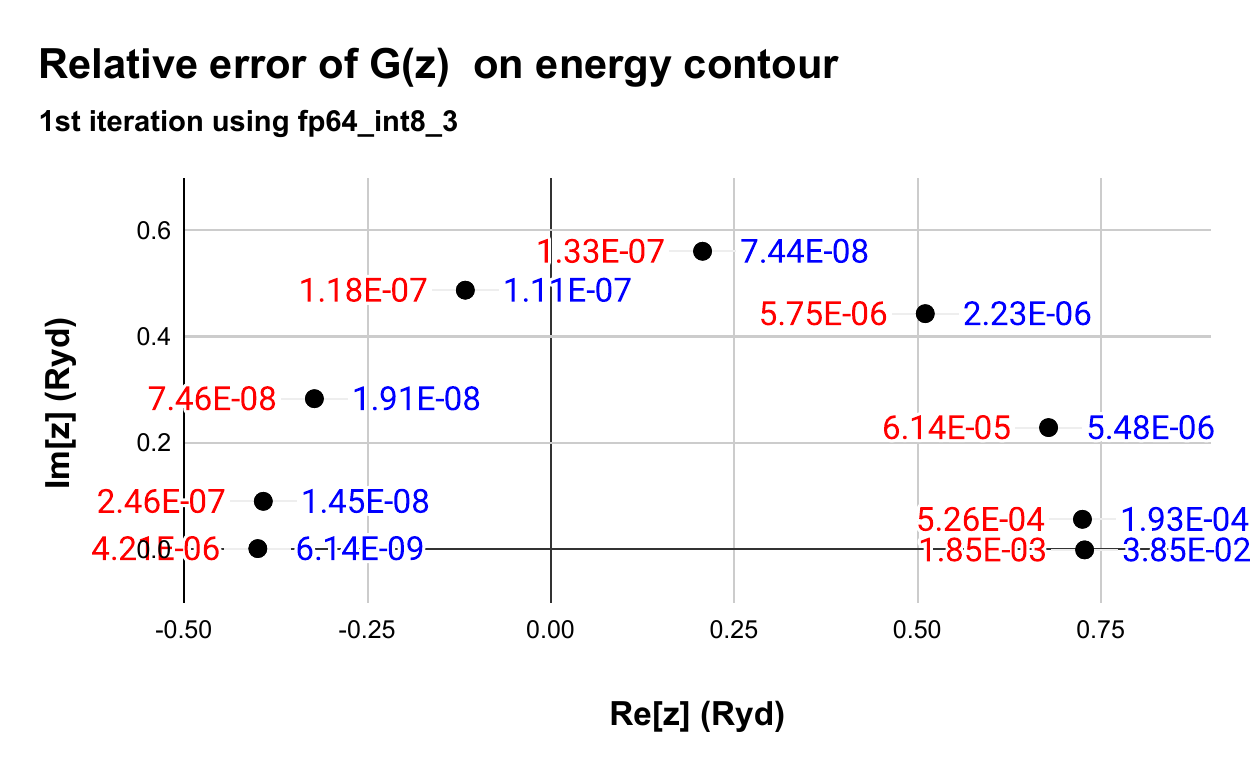}
  \end{subfigure}
  \begin{subfigure}[b]{0.42\textwidth}
    \includegraphics[width=\linewidth]{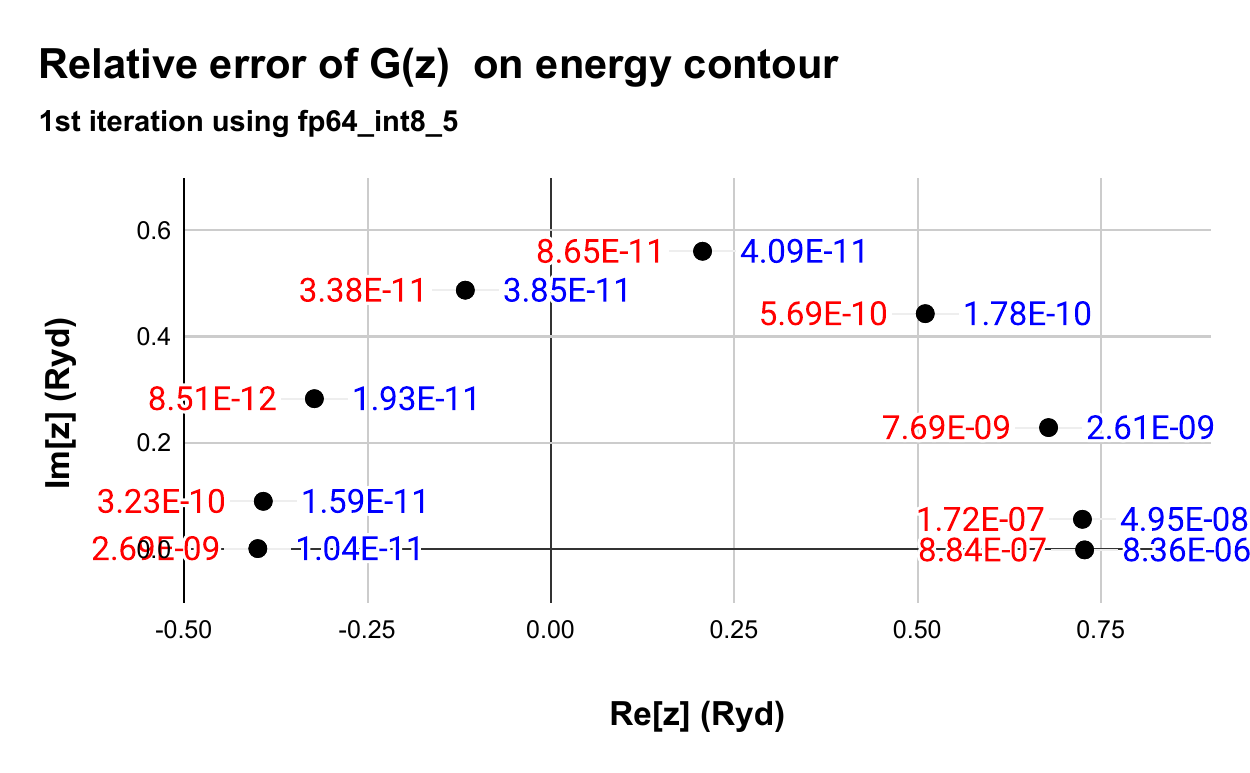}
  \end{subfigure}
  \vspace{-5.5mm}
  \caption{
    Relative error of real(blue) and imaginary(red) parts of $G(z)$ on energy contour(black dots) from 1st iteration using fp64\_int8\_3 and fp64\_int8\_5
  }
  \label{fig:max}
\end{figure}

When analyzing the $G(z)$ data at each energy point, an interesting pattern of max\_real and max\_imag invisible in FP64 mode is observed across split numbers from 3 to 6. These errors peak in an isolated region near the Fermi energy (0.72 Ryd on the real axis), as shown in Figure \ref{fig:max} for split numbers 3 and 5. As energy points (black dots) move counterclockwise along the contour $z$ away from this region, relative errors in Re[$G(z)$] (blue) and Im[$G(z)$] (red) decrease exponentially, with split number 3 showing greater sensitivity due to its lower tolerance for ill-conditioned objects. This behavior arises from physical states near this region, and the $G(z)$ has poles on those states. Our results above represent the first MuST runs with tunable precision to quantify this pattern. This raises a longstanding yet underexplored question: can the tunable precision approach generally quantify and separate the ill- and well-conditioned domains and determine what necessary precision for each? Given ozIMMU’s performance drops quadratically with increasing split numbers, minimizing splits while maintaining accuracy is critical. Particularly for MuST, where errors originate from an isolated region, regularizing the iteration kernel or dynamically adjusting the split number in that region offers a promising approach to improve accuracy with fewer splits.

Performance is not our focus here, as the GH200 architecture’s INT8 performance (1,979 TOPS) does not significantly outpace its FP64 performance (67 TFLOPS). Consequently, our MuST run with split number 6 takes 731.799 s, which is much slower than the 412.149 s using native FP64. This is evident in the DGEMM benchmarks \cite{uchino_performance_2025} we ran with a 2048x2048 matrix, the typical size in MuST, where split number 6 achieves 20.35 TFLOPS versus FP64’s 62.52 TFLOPS. However, next-generation AI hardware, like the GB200 with projected 5,000 TOPS of INT8 and 40 TFLOPS of FP64, is expected to deliver higher IMMU emulation performance compared to FP64.

\section{Conclusion}

In this study, we presented a preliminary investigation into emulating floating-point matrix multiplication operations using INT8 data types. This approach, combined with our previously developed automatic BLAS GPU offloading tool, is applied to the well-known FP64-based quantum physics code, LSMS in the MuST suite. 
Using the Ozaki Scheme, which splits floating-point numbers into multiple INT8 variables, 
we explored various splitting configurations to identify the optimal balance between efficiency and the precision necessary for scientific simulations. 
Our results demonstrate that this emulation scheme is both effective and accurate for FP64-based scientific applications, enabling simulations to be carried out with a reduced bit-width for floating-point numbers without sacrificing accuracy.

Our emulation approach offers a broader and generic solution for scientific simulations, distinguishing itself from the typical mixed-precision studies, where specific solver algorithms are modified to accommodate lower precision in intermediate computations. 
Unlike those approaches, our method preserves the integrity of the original algorithm and code without modification while optimizing the utilization of hardware resources. 
We advocate for a re-evaluation of precision requirements in scientific applications and determine the precision or the number of significant bits that are truly necessary. 
We encourage the adoption of low-bitwidth types to enhance hardware utilization, particularly in the context of the growing influence of AI on new hardware. 
Additionally, we urge collaboration between hardware developers and computational scientists to design optimal data types that can better meet the demands of future scientific computing along with AI.
\begin{acks}
This work is supported by the National Science Foundation through the OAC-2402542 award. It is also supported in part by the OAC-2139536 award. The authors thank Dr. Yang Wang from PSC for insightful discussions on the implementation details of MuST. The authors also acknowledge AI tools including Grok, ChatGPT, and Claude that helped revise the manuscript. 
\end{acks}

\bibliographystyle{ACM-Reference-Format}
\bibliography{reference}

\end{document}